# Evaluation and Ensembling of Methods for Reverse Engineering of Brain Connectivity from Imaging Data

Bisakha Ray[1], Alexander V. Alekseyenko[2], Sisi Ma[1], Alexander Statnikov[1], Constantin Aliferis [5, 6]

[1]Center for Health Informatics and Bioinformatics, New York University School of Medicine, New York, [2]Department of Public Health Sciences, Department of Oral Health Sciences, Biomedical Informatics Center, Medical University of South Carolina, South Carolina, [5] Department of Biostatistics, Vanderbilt University, Tennessee, [6] Institute for Health Informatics, University of Minnesota, Minnesota, USA

**Key terms:** Brain Connectomics, Brain Modeling, Neuroimaging, Network Reconstruction

*Corresponding author:*

Bisakha Ray

Center for Health Informatics and Bioinformatics, New York University Langone Medical Center,

227 East 30th Street, 757 E,

New York, NY: 10017, USA

Phone: +1-662-715-9326

E-mail: bisakha.ray@nyumc.org

# ABSTRACT

Brain science is an evolving research area inviting great enthusiasm with its potential for providing insights and thereby, preventing, and treating multiple neuronal disorders affecting millions of patients. Discovery of relationships, such as brain connectivity, is a major goal in basic, translational, and clinical science. Algorithms for causal discovery are used in diverse fields for tackling problems similar to the task of reconstruction of neuronal brain connectivity. Our aim is to understand the strengths and limitations of these methods, measure performance and its determinants, and provide insights to enhance their performance and applicability. We performed extensive empirical testing and benchmarking of reconstruction performance of several state-of-the-art algorithms along with several ensemble techniques used to combine them. Our experiments used a clear and broadly relevant gold standard based on calcium fluorescence time series recordings of thousands of neurons sampled from a previously validated realistic, neuronal model. Correlation, entropy-based measures, Cross-Correlation for short time lags, and Generalized Transfer Entropy had the best performances with area under ROC curve (AUC) in the range of 0.7–0.8 even for smaller sample sizes of n = 100 to 1,000 and converged quickly (at less than n = 1,000). Ensembles of best-performing methods using random forests and neural networks generated AUC of ~0.9 with n = 10,000. Several important insights regarding parameter choice and sample size were gained for guiding the experimental design of studies. Our data are also supportive of the feasibility of reliably reconstructing complex neuronal connectivity using existing techniques.

# ACRONYMS

Local field potentials: LFP

Functional magnetic resonance imaging: fMRI

Electroencephalography: EEG

Multivariate vector auto-regressive: MVAR

Granger causality: GC

Blood oxygenation level-dependent: BOLD

Transfer entropy: TE

Generalized Transfer Entropy: GTE

Clustering coefficient: CC

Neural Simulation Tool: NEST

Area under ROC curve: AUC

Area under Precision-Recall curve: AUPR

Support Vector Machines: SVMs

Artificial Neural networks: ANNs

Random Forests: RFs

Receiver Operator Characteristic: ROC

Confidence interval: CI

**Introduction**

The launch of international initiatives such as the BRAIN Project (Shepherd *et al.*, 1998, Padmanabhan *et al.*, 2015), and the Human Brain Initiative (Insel *et al.*, 2013) has generated intense interest in deciphering the complicated circuitry of the human brain. Reverse engineering of networks has been a commonly tackled problem in genomics and systems biology but is now gaining momentum in neuroscience.

Unraveling the brain structure at the neuronal level at a large scale is an important step in brain neuroscience. It promises to have tremendous impact for gaining insight into animal and human intelligence and learning capabilities and in understanding and treating neuronal diseases and injuries. However, understanding the brain's anatomical structure using conventional methods by disentangling the complicated intertwining of the dendrons and axons is extremely difficult and time intensive. Another way to attempt to solve this problem is to reconstruct the connections of a neuronal network from observational time series data of neuronal activity of thousands of neurons. This temporal, observational data can be obtained with state-of-the-art calcium fluorescence imaging (Guyon *et al.*, 2014).

Calcium fluorescence activity recordings of neuronal cultures provide observational time-series data of neurons *in vivo*. It allows for the recording of activity of thousands of neurons non-invasively. The cells to be monitored are preloaded with calcium-sensitive dye. In case of neuronal activity, there is an influx of calcium into the cells related to the activity. This calcium binds to the dye and generates fluorescence. This experimental model is less time consuming and tedious than other conventional, extensive experimental techniques such as axonal tracing that involve physically injecting tracers to identify connections between neurons (Guyon et al., 2014).

Several other modalities like *local field potentials* (LFP), *functional magnetic resonance imaging* (fMRI), *electroencephalography* (EEG), etc. have been used in neuroscience for causal network reconstruction. Calcium fluorescence imaging data is a next-generation assay that has the advantages of being non-interventional, non-invasive, and inexpensive.

Very few studies have exploited the benefits of observational experiments from neuronal cultures, which contain in their microcosms the self-organizing activity of neurons.

Therefore, our study focuses on calcium fluorescence imaging data.

*Causal Structure Learning in Longitudinal Non-Experimental Data in Neuroscience*

When randomized experimental designs are infeasible and not practical, researchers must resort to the use of observational or non-experimental data. Lack of randomization of values of causes entails that unmeasured confounders potentially bias estimated effects of those causal variables. Hence, the challenge of estimating causal effects with observational data can be incredibly difficult (Winship & Morgan, 1999). However, observational data have been used to uncover causal effects in realistic datasets with available gold standards (Maathuis *et al.*, 2010). These predictions could again be validated or followed with interventions in experiments and can never replace experimental studies. We next review the approaches taken at measuring (causal) effects; we will formally define them in a later section of this paper.

<u>Correlation-based methods</u>
Correlation, cross-correlation, and partial correlation have been used as graph-theoretical measures in measuring brain connectivity in distinct cortical areas (Bullmore & Sporns, 2009,

Sporns *et al.*, 2005). Correlation between structural and functional connectivity have been done in some studies using resting state fMRI (Hagmann *et al.*, 2010). In neuroscience, Pearson's correlation coefficient has been used to determine functional connectivity between pairs of variables. Partial correlation has been estimated to identify even finer associations between behavior and brain function (Lynall *et al.*, 2010). Correlation has been used to study brain connectivity and neuronal diseases like schizophrenia and Alzheimer's (Padmanabhan et al., 2015, Lynall et al., 2010, He *et al.*, 2008). (Bassett & Bullmore, 2006) suggested that apart from using measures such as partial correlation to identify functional connectivity between regions, one can also use frequency-domain based measures like correlation wavelets or Fourier transforms.

Entropy-based methods

Neuronal activity is multivariate and stochastic in nature. One can analyze such activity using information theoretic measures. Information theoretic-based measures such as mutual information have been used to identify structural and functional connectivity in the brain (Hagmann et al., 2010, Bullmore & Sporns, 2009, Lynall et al., 2010). Attempts to combine such measures to identify distinct neural subunits and their functional and structural interplay have been made (Honey *et al.*, 2007, Sporns *et al.*, 2000). Several variations of mutual information like cross mutual information have been used to study EEG in patients with Alzheimer's disease (Jeong *et al.*, 2001).

Granger Causality-based methods

As Granger Causality (GC) is based on temporal causal precedence, it has been beneficial for dynamic modeling and direct mapping of influences in brain connectivity (Roebroeck *et al.*, 2011, Roebroeck *et al.*, 2005). *Multivariate vector auto-regressive* (MVAR) GC techniques have been applied to brain imaging data to identify how preceding neural activity triggers subsequent disorders (Hamilton *et al.*, 2011). Several other techniques such as using the summary time series instead of the raw time series and combining MVAR GC, down sampling of the time series, and graph theoretic concepts were successful in investigating causal brain networks (Deshpande *et al.*, 2009). Not only have GC been useful in predicting directionality, it can also be used to predict *blood oxygenation level-dependent* (BOLD) activity levels (Jiao *et al.*, 2011). However, naïve inference of GC over fMRI signals as a measure of effective connectivity between neuronal populations can be misleading. While linear estimation of GC can provide initial good estimates, in order to capture more complex causal relationships, nonlinearity has to be introduced (Sitnikova *et al.*, 2008). In another study, realistic BOLD time series corresponding to fMRI data was generated (Smith *et al.*, 2011). The simulations were based on Dynamic Causal Modeling model (Friston *et al.*, 2003) fMRI forward model. Different reconstruction approaches tested by the authors were correlation, partial correlation, mutual information, variations of GC such as conditional GC, pairwise GC, directed GC, and Geweke's GC, partial Directed Coherence, conditional dependence measures, and Bayes net methods. Metrics included connection strengths and z-scores. While partial correlation and Bayes net performed well, sensitivity of lag-based methods such as GC was poor.

As linear versions of auto-regressive models for Granger causality may not be enough to capture complex neuronal signals, Transfer Entropy which is non-linear may be utilized. For Gaussian variables, Transfer Entropy reduces to Granger causality (Barnett *et al.*, 2009)

<u>Transfer Entropy-based methods</u>
Transfer entropy metric (Schreiber, 2000), unlike GC, is not based on a prior model and can accommodate non-linearity (Vicente *et al.*, 2011). This metric is able to capture directed causal interactions to facilitate inference of driver and response variables. In several studies, transfer entropy was used to reconstruct interactions between different brain regions (Honey et al., 2007). In paper by Chávez *et al.* (2003), transfer entropy was used to deal with hidden confounders and common history between two variables in EEG data. Transfer entropy does not assume a specific underlying model of interaction of the variables (Vicente et al., 2011).

**Materials and Methods**

In our study, the data for reconstruction consists of simulated calcium fluorescence imaging recordings of the activity of neurons in brain networks. The aim is to predict the connections of the neuronal network from the calcium fluorescence imaging time series data. There is no available ground truth or known connections between neurons for experimental data. Hence, we do not have a way to validate the inferred network from experimental data. We use simulated data with 'surrogate' ground truth data from a broadly accepted ground truth model with behavior similar to realistic neuronal networks coming from the Connectomics challenge (Kaggle, 2014). The fluorescence amplitude signal distributions can be characterized by a region of low fluorescence activity with noise, followed by a region of intermediate firing rate (which provide best reconstruction), followed by regions of highly synchronized activity or network bursts. The low fluorescence noise-dominated region assumes a Gaussian-like shape. The high fluorescence region has a long tail (Stetter *et al.*, 2012).

The calcium fluorescence signals were simulated by treating the spiking dynamics. This was done using a model (Vogelstein *et al.*, 2009) that gives rise to an initial rapid increase in fluorescence due to activation followed by a slow decay. This corresponds to the influx of calcium into cells after firing and binding of calcium to fluorescence probes. 180,000 samples (1 hour at 50 frames per second) were generated.

For each pair of neurons, the problem we consider is to determine whether a connection exists between them.

The entries in a matrix of real interactions is as shown in Table 1. The presence or absence of connections is indicated by means of discrete values of 1 and -1, respectively. The entries in a simulated network matrix will be similar to the ground truth topology of a real network having 1 and -1 representing the presence or absence of connections. An example interaction matrix derived from the calcium fluorescence signals of four neurons A, B, C, and D using Pearson's correlation as a network reconstruction technique would be as shown in Table 2. We ignore self-connections. Here scores, and not discrete connections, on edges between pairs of neurons are obtained.

*Causal Structure Learning Techniques*

Several reconstruction methods are used for causal structure learning. The reconstruction strategies are correlation-based, entropy-based, and GC-based.

Correlation-Based Measures

We use two correlation-based measures—Pearson's correlation and Cross-correlation—for reconstruction.

Pearson's Correlation

*Pearson's correlation coefficient* is a measure of the degree of the linear association between variables.

For variables $X$ and $Y$ with means $\bar{X}$ and $\bar{Y}$, the Pearson's correlation, for $X$ and $Y$ is calculated according to (1)

**TABLE 1.** Matrix showing presence or absence of real neuronal connections.

|   | A  | B  | C  | D  |
|---|----|----|----|----|
| A | -1 | 1  | 1  | -1 |
| B | -1 | -1 | -1 | -1 |
| C | 1  | -1 | -1 | 1  |
| D | 1  | 1  | 1  | -1 |

**TABLE 2.** Inferred matrix showing scores on connections from reconstruction method.

|   | A    | B    | C    | D    |
|---|------|------|------|------|
| A | 0    | 0.68 | 0.95 | 0.72 |
| B | 0.68 | 0    | 0.56 | 0.83 |
| C | 0.95 | 0.56 | 0    | 0.41 |
| D | 0.72 | 0.51 | 0.41 | 0    |

$$\text{Pearson's correlation}(X,Y) = \frac{\sum_{i=1}^{n}(X_i-\bar{X})(Y_i-\bar{Y})}{\sqrt{\sum_{i=1}^{n}(X_i-\bar{X})^2}\sqrt{\sum_{i=1}^{n}(Y_i-\bar{Y})^2}}, \quad (1)$$

where $n$ is the number of dimensions of $X$ and $Y$. Methods based on this measure ignore the temporal nature of the data and treat each observation as either independent or exchangeable. This simplifying assumption becomes less valid as the sampling distribution diverges from stationary.

## Cross-correlation

*Cross-correlation* involves correlating variables over many different time lags. The shifting correlation in Cross-correlation can be achieved mathematically by introducing a time lag. The Cross-correlation, at time lag $d$ between two time series $X$ and $Y$, with means $\bar{X}$ and $\bar{Y}$ can be calculated

$$\text{Cross-correlation}(X,Y) = \frac{\sum(X_i-\bar{X})(Y_{i-d}-\bar{Y})}{\sqrt{\sum(X_i-\bar{X})^2}\sqrt{\sum(Y_i-\bar{Y})^2}}. \quad (2)$$

Cross-correlation (2) can be computed for all time lags $d = 0, 1, 2,\ldots$ At time lag 0, this reduces to calculating Pearson correlation coefficient (1) between time series $X$ and $Y$. Cross-correlation assigns a score to each possible link between two nodes based on the highest cross-correlogram peak for assigned time lags.

## Entropy-Based Measures

Entropy of a discrete random variable $X$ defined on a probability space $p$ is defined as in (3)

$$H(p) = H(X) = -\sum_{x \in X} p(x) \log_2 p(x), \quad (3)$$

where $p(x)$ is the probability that the variable $X$ assumes value $x$. The joint entropy of two variables $x$ and $y$ is defined as in (4),

$$H(X,Y) = -\sum_{x \in X} p(x,y) \sum_{y \in Y} p(x,y) \log_2 p(X,Y). \quad (4)$$

Entropy is the amount of information change in a random variable or the average uncertainty of a random variable.

For example, let us assume we have a coin with heads on both sides. The outcome of every toss is determined and there is no uncertainty in the outcome. In such a case, the entropy or information change is zero. However, if we have a fair coin, the entropy of outcome of a toss is maximized as it is highly uncertain.

## Gini Index

The Gini Index measures the impurity in D, a set of training tuples. The m classes in (5) correspond to the unique values or states in the vector D which is a discretized calcium fluorescence time series,

$$Gini(D) = 1 - \sum_{i=1}^{m} p_i^2 , \quad (5)$$

where, $p_i$ is the probability that a tuple in $D$ belongs to class $C_j$ and is estimated by $|C_{i,D}|/|D|$.

The sum is calculated over *m* classes (Han *et al.*, 2011).

Let us assume F = {0, 1, 1, 2, 0, 1} represents a discretized fluorescence time series. The total length of the series here is 6. The number of unique values or classes here is 3 (0, 1, and 2). 0 occurs in 2 out of 6 positions, 1 in 3 out of 6 positions, and 2 in 1 out of 6 positions. So, the Gini index here can be calculated as $1 - [(2/6)^2 + (3/6)^2 + (1/6)^2]$.

When we choose discretization by looking at variations from a baseline to define states our choice of classes correspond to isolated potential spiking events which maximize impurity.

## Mutual Information

If two random variables $X$ and $Y$ have probabilities $P(X)$ and $P(Y)$, then their mutual information $I(X,Y)$ is defined as in (6) (Church & Hanks, 1990),

$$I(X;Y) = H(X) - H(X|Y) = H(Y) - H(Y|X) = \sum_{x,y} p(x,y) \log \frac{p(x,y)}{p(x)p(y)}, \qquad (6)$$

where $p(x, y)$ is the joint probability distribution function of variables $X$ and $Y$, $p(.)$, in abuse of notation, designates either the marginal density functions of $p_X(x)$ or $p_Y(y)$ of $X$ and $Y$, respectively, as clear from the argument.

For two discrete random variables X and Y, mutual information compares the values of their joint probability $p_{XY}(x, y)$ with the product of the values of marginal probabilities $p_X(x)$ and $p_Y(Y)$. Mutual information compares the probability of observing X and Y together versus the probabilities of observing X and Y independently.

Generalized Transfer Entropy

In its original formulation, the *transfer entropy* (TE) (Schreiber, 2000), from $Y$ to $X$ for two Markov processes $X$ and $Y$ of order $k$, is defined as in (7),

$$TE_{Y \to X} = \sum P\left(x_{n+1}, x_n^{(k)}, y_n^{(k)}\right) \log \frac{P(x_{n+1}|x_n^{(k)}, y_{n+1}^{(k)})}{P(x_{n+1}|x_n^{(k)})}, \quad (7)$$

where $x_n^{(k)}$ is a vector of length $k$ whose entries are samples of $X$ at time $n, n-1, \ldots, 1$. If $D_{x,t}$ represents the calcium fluorescence time series for neuron $x$ at time $t$, then $x_n = D_{x,n+1} - D_{x,n}$. TE can be interpreted as the distance between the probability distributions or the *Kulback–Leibler distance* between the single-node transition matrix $P(x_{n+1}|x_n^{(k)})$ and the two-node transition matrix $P\left(x_{n+1}, x_n^{(k)}, y_n^{(k)}\right)$. TE is zero if the two probability distributions are identical or the distance between them is zero. TE greater than zero indicates dependence of $x$ on past values of $y$ (Stetter et al., 2012).

In the formulation of Generalized Transfer Entropy (GTE) (Stetter et al., 2012), , the model postulates that network switches between different dynamical states of high synchronous activity or bursting and asynchronous activity or non-bursting. Representative fluorescence histograms of the two networks are shown in Figure 1(a) and Figure 1(b).

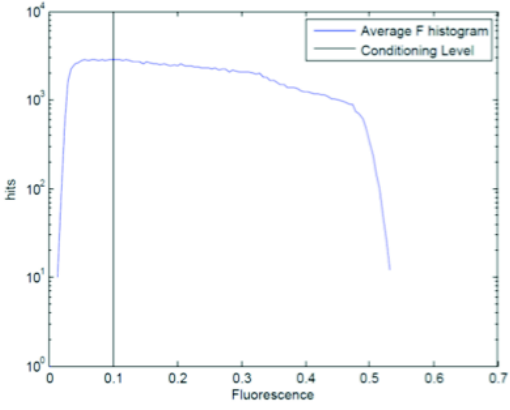 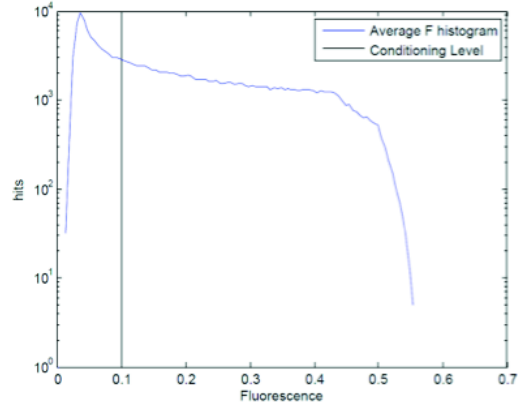

**FIG. 1(a).** Calcium fluorescence histogram of Network 1.

**FIG. 1(b).** Calcium fluorescence histograms of Network 2.

The signal is averaged for the whole fluorescence time series as shown in (8) to restrict the evaluation of the network to a consistent range reflective of the averaged global dynamics (Stetter et al., 2012).

$$g_t = \frac{1}{N}\sum_{i=1}^{N} x_i^{(t)}. \qquad (8)$$

Only data points in which the fluorescence $g_t$ is below this threshold called a conditioning level or $\tilde{g}$ are included in the analysis. Hence, TE using the above two considerations is reformulated as in (Stetter et al., 2012) (9)

$$TE^*_{Y \to X}(\tilde{g}) = \sum P\left(x_{n+1}, x_n^{(k)}, y_{n+1}^{(k)} | g_{n+1} < \tilde{g}\right) \log \frac{P(x_{n+1}|x_n^{(k)}, y_{n+1}^{(k)}, g_{n+1} < \tilde{g})}{P(x_{n+1}|x_n^{(k)}, g_{n+1} < \tilde{g})}. \qquad (9)$$

Granger Causality-Based Method

From (Granger, 1980), by definition, $Y$ *Granger-causes* $X$ if for some $s > 0$, the mean squared error of a forecast of $X_{t+s}$ based on $(X_t, X_{t-1}, \ldots)$ is larger than based on $(Y_t, Y_{t-1}, \ldots)$ and ($X_t$, $X_{t-1}$, …). Conversely, $Y$ fails to Granger-cause $X$ if for all $s > 0$, the mean squared error of a forecast of $X_{t+s}$ based on ($X_t$, $X_{t-1}$, …) is the same as that is based on ($Y_t$, $Y_{t-1}$, …) and ($X_t$, $X_{t-1}$, …).

The intuition behind these formulas is that a simple test can differentiate between direct (or proximal) causal influences from remote causal influences and confounded (non-causal) correlations.

*Experimental Design*

Datasets and data preparation

The datasets generated for the experiments used a realistic neuronal connectivity network simulator (Stetter et al., 2012, Guyon et al., 2014). Training and test data for networks of around 100–500 neurons are generated using the neuronal connectivity simulator. Table 3 provides a description of the generated networks. The connection and inhibitory probability are based on similar studies from literature (Stetter et al., 2012).

The network structure is as close as possible to real anatomical topologies and dynamical behaviors. Connectivity models taking into account realistic neuronal proximity and node clustering is used. In the network construction phase, the simulator generates nodes randomly. The connection probability between nodes is ~0.12 which is close to the sparse connectivity observed in actual cortical circuits. To generate connections, an initial connectivity matrix with uniform connection probability is created. A random pair of neurons is selected and their connections are crossed while maintaining the *indegree* and *outdegree* of the nodes (Stetter et al., 2012). The clustering coefficient (CC) of a network by Watts and Strogatz (Watts & Strogatz, 1998) is defined as follows. Let us suppose a vertex V in a graph has k neighbors. If every vertex is connected to every other vertex in the graph then a maximum of $\binom{k}{2} = \frac{k \times (k-1)}{2}$ edges can exist between them. Let $C'$ be the fraction of these possible edges that actually exist between a pair of nodes. The CC is then the average of $C'$ overall V.

**TABLE 3.** Description of datasets.

| Network Identifier | Number of nodes | Connection Probability | Inhibitory Probability | Clustering Coefficient | Indegree Median [95% CI] | Outdegree Median [95% CI] |
|---|---|---|---|---|---|---|
| Network 1 | 100 | 0.12 | 0.2 | 0.6 | 12 [7, 18] | 11 [5, 20] |
| Network 2 | 100 | 0.16 | 0.2 | 0.6 | 16 [10, 25] | 16 [9, 24] |
| Network 3 | 500 | 0.12 | 0.2 | 0.16 | 59 [46, 74] | 59 [47, 74] |
| Network 4 | 500 | 0.16 | 0.2 | 0.18 | 79 [64, 97] | 79 [65, 96] |

The crossing process of the connections was repeated till a desired CC was obtained. The diagonal entries of the connectivity matrix representing a neuron's connections to itself are removed.

For modeling, an excitatory *leaky integrate-and-fire model* (Liu & Wang, 2001) of spiking neurons is used. The dynamics of the networks mimicking the spontaneous firing of neuronal cultures were simulated using the Python NEST simulator (Brette *et al.*, 2007). The activity of neuronal cultures is generally characterized by the occurrence of irregular switching between states of asynchronous and highly synchronous activity known as network bursts. A clear bursting regime is defined for the networks (Stetter, 2012).

For the fluorescence model, calcium fluorescence time series is simulated from the network using a model described in (Wen *et al.*, 2009) which takes into account time averaging and light scattering effects. This model describes the activity of the intracellular concentration of calcium by an initial rapid increase in the concentration due to activation followed by a decrease.

The signals contain considerable noise not just from light scattering but also from calcium fluctuations independent of spiking activity, calcium fluctuations from nearby cells, noise during image acquisition etc. During action potential, there is a sudden influx of calcium into the cells. This is observed as a sharp rise in the calcium fluorescence traces. The low frequencies correspond to decay in the fluorescence. A study of the raster properties of the data such as inter-burst interval shows that the signals generated closely resemble distributions of real data.

The total number of samples correspond to realistic length of recordings of 50 frames per second in an hour or, 50 *60 * 60 total number of samples = 180,000 samples. The calcium fluorescence time series signals are continuous. However, for network reconstruction they have been discretized using binning. If $D_{x,t}$ represents the calcium fluorescence time series for neuron $x$ at time $t$, then $x_n = D_{x,n+1} - D_{x,n}$. In order to analyze the performance of the algorithms and benchmark them, we generated calcium fluorescence time series from the simulated neural network.

<u>Study design</u>

Algorithms for Network Reconstruction

As shown in Figure 1 (a) and Figure 1 (b),

The calcium fluorescence intensity histogram has a characteristic right-skewed shape which corresponds to switching between bursting and non-bursting states. The high fluorescence region has a long-tail. The distribution of intensities in the low fluorescence region has a Gaussian-like shape. The mean of the fluorescence activity for each neuron was used. So, the histogram corresponds to a histogram of average fluorescence signals.

Reconstruction quality is heavily affected when we focus on the fluorescence region of intermediate firing rate above the Gaussian in the histograms in Figure 1 (a) and Figure 1 (b). The selection of this region corresponds to sampling the data points most representative of the network dynamics and excludes bursts. This is essential to achieving good signal to noise ratio. As a preprocessing step, very basic discretization or binning is performed to filter out the spiking events and improve the signal-to-noise ratio.

The same preprocessing involving discretization and conditioning with respect to neuronal dynamics is applied for all analyses. For reconstruction, as shown in Table 4, we considered correlation based, information gain, GC, and GTE algorithms. All of these algorithms have been used in prior studies for reconstruction (Garofalo *et al.*, 2009, Singh & Lesica, 2010, Ostwald & Bagshaw, 2011, Biffi *et al.*, 2011). Cross-Correlation, Mutual Information, GC, and GTE methods are parametric methods as the results obtained depend on the choice of the bin size or temporal window slice. The synaptic time intervals of interactions of neurons are much shorter than the time for recording. In GTE, we have considered instantaneous causal interactions in terms of time-bin units. To account for a slightly slower time scale of interaction, a Markov order of 2 was considered. For Cross-Correlation and Granger, we have considered both extremely short lags of 1 and a slower one of 10. The Cross-Correlation or Granger lags time here corresponds to whether a neuron in a particular region fires on an average earlier or later than another neuron in the networks.

**Table 4:** Algorithms and parameters [*N1 = Number of parameterization for obtaining likelihood of edges in the graph (some scores, e.g. pairwise correlations), N2 = Number of parameterization for obtaining specific graphs.*]

| Causal Discovery Approach | Core Algorithm [N1\|N2] | Parameters |
|---|---|---|
| Granger Causality based | Granger causality [2\|4] | time lag = 1, conditioning level = 0.10 |
| | | time lag = 1, conditioning level = 0.15 |
| | | time lag = 10, conditioning level = 0.10 |
| | | time lag = 10, conditioning level = 0.15 |
| Other | Mutual Information [1\|2] | conditioning level = 0.10 |
| | | conditioning level = 0.15 |
| | Gini Coefficient [1\|2] | conditioning level = 0.10 |
| | | conditioning level = 0.15 |
| | Pearson's Correlation [1\|2] | conditioning level = 0.10 |
| | | conditioning level = 0.15 |
| | Cross-Correlation [2\|4] | time lag = 1, conditioning level = 0.10 |
| | | time lag = 10, conditioning level = 0.10 |
| | | time lag = 1, conditioning level = 0.15 |
| | | time lag = 10, conditioning level = 0.15 |
| | GTE [1\|2] | Conditioning level = 0.10, Number of bins = 3, Markov order of the process = 2 |
| | | Conditioning level = 0.20, Number of bins = 3, Markov order of the process = 2 |

We subsampled the simulated data to generate samples of sizes 500, 1,000, 10,000, and 100,000 and for 1,000, 500, 100, and 10 repetitions for conditioning levels of 0.10 and 0.15. Results were averaged over the number of runs for a given network and sample size. An important feature of this data is the temporal structure, which must be preserved by the sampling protocol. Hence, for Cross-Correlation and Granger causality, which utilize temporal constraints in the form of time lags, we subsampled from random chunks of sizes 500, 1,000, 10,000, and 100,000 in the original time series. The random chunks would have consecutive samples. For Correlation, Mutual Information, Gini, and GTE, which did not require a temporal constraint, we generated samples randomly from the original time series. Interpretation of our results is conditioned on respective sampling for each class of methods.

A benchmark comparison between the reconstruction performances of the procedures based on sample size was performed. *Area under ROC curve* (AUC) and *Area under Precision-Recall curve* (AUPR) with 95% *confidence interval* were averaged for all runs for all networks for a particular sample size.

Ensemble learning

Ensemble learning or learning on meta- features generated by weighing base predictors can improve classification (Wang *et al.*, 2009, Kim *et al.*, 2003, Wang *et al.*, 2003). Based on our benchmarking study, in phase 2 of the design we generated an ensemble of features from the scores of best-performing algorithms. The features were constructed by concatenating scores from correlation, cross-correlation for time lag 1, mutual information, Gini index, and GTE. The machine learning-based classification methods used were Support Vector Machines (SVMs)(Burges, 1998), Artificial Neural networks (ANNs) (Hornik *et al.*, 1989, Haykin &

Network, 2004), and Random Forests (RFs)(Breiman, 2001). Our algorithmic approach is outlined in Algorithm.

**ALGORITHM.** Procedure to generate connections from ensemble learners.

> 1. Generate simulated time series data with realistic properties for each network.
> 2. Repeat 3-5 for 1000, 500, 100, and 10 runs respectively.
> 3. Draw random subsamples of sizes 500, 1000, 10000, and 18000 from original datasets with replacement.
> 4. For each sample size:
>    i. Apply discretization and conditioning.
>    ii. Generate connectivity scores using reconstruction algorithms—Correlation, Cross-Correlation with time lag 1, Mutual Information, Gini coefficient, and GTE.
>    iii. Concatenate scores from ii to generate features for connections between each pair of neurons. Training labels are obtained from ground truth data indicating presence or absence of connections.
>    iv. Perform 10-fold cross-validation repeated over 10 iterations using classifiers RF, linear SVM, and NNs.
>    v. Generate AUC and AUPR.
> 5. Calculate mean AUC and AUPR with 95% C.I. for each sample size for each network.

Linear SVMs are supervised classification algorithms that classify samples into two classes, here the presence or absence of synaptic connections between a pair of neurons, by calculating the maximal-margin hyperplane separating them. We have used a LIBSVM (Chang & Lin, 2011) MATLAB interface with a linear SVM and a cost parameter of 1.

RFs are an ensemble classification method, which uses bagging and constructs multiple decision trees on subsamples with replacement from the original dataset at training time. The label is the mode of the classes of individual decision trees. We have used an R implementation of the RF (Liaw & Wiener, 2002) with 100 decision trees and no pruning. We have also performed permutation testing to robustly test for overfitting, which revealed that results from RF were significantly better than random AUC values.

ANNs are pattern recognition tools mimicking biological neural networks. Given a specific prediction task, NNs use the training input patterns to learn the class of functions by minimizing error. The connections between neurons have weights which are fine-tuned while learning. We have used the MATLAB Neural Network Toolbox (Demuth & Beale, 1993) with 1 hidden layer, 10 hidden neurons, and 1000 training epochs.

All experiments were performed on CentOS 6.3 with MATLAB v2013a, Python 2.7.3 and R (2013a).

Performance and Error Estimation

We used repeated nested n-fold cross validation with n=10 and 10 repeats (Hsu *et al.*, 2003) and averaged results over the 10 repetitions for the ensembles generated from random subsampling of the original calcium fluorescence data. The cross-validation procedure divides the subsamples drawn into 10 non-overlapping balanced subsets. The process is then repeated 10 times with 9 sets used for training and 1 for testing. We have also robustly tested for overfitting by performing permutation testing.

Performance Metrics

The network reconstruction performance was evaluated with the AUC and AUPR (Narendra *et al.*, 2011).

The AUC is equivalent to the area under the curve obtained by plotting *sensitivity* or true positive rate against 1-*specificity* or false positive rate at different thresholds. It represents the probability that the classifier will rank a randomly chosen instance from the positive class higher than a randomly chosen instance of the negative class.

Area under Precision Recall curve (AUPR) is obtained by plotting precision and recall at every position in the ranked sequence of samples.

Precision or positive predictive value is the fraction of retrieved positive instances that are relevant. Recall or sensitivity is the fraction of relevant instances that are retrieved. AUPR represents the average of precision across all recall values.

We rank the scores on edges returned by the reconstruction method. AUPR would represent the probability that if an edge existing in the "gold standard" is selected from this ranked list, an edge above it in the ranked list of edges will also be an edge existing in the "gold standard".

**Results**

In our benchmarking study, for random chunks of sample sizes 500, 1,000, 10,000, and 100,000, the network reconstruction performance was evaluated for 1000, 500, 100, and 10 runs for conditioning levels of 0.10 and 0.15. Results were averaged over all the runs for a given network and random chunk size. As can be observed in Figure 2 from the *Receiver Operator Characteristic* (ROC), the sample size required for reconstruction at maximal performance is relatively small, not exceeding 1000 time samples. Correlation-based and Entropy-based techniques (Gini and Mutual Information) had the best reconstruction performances.

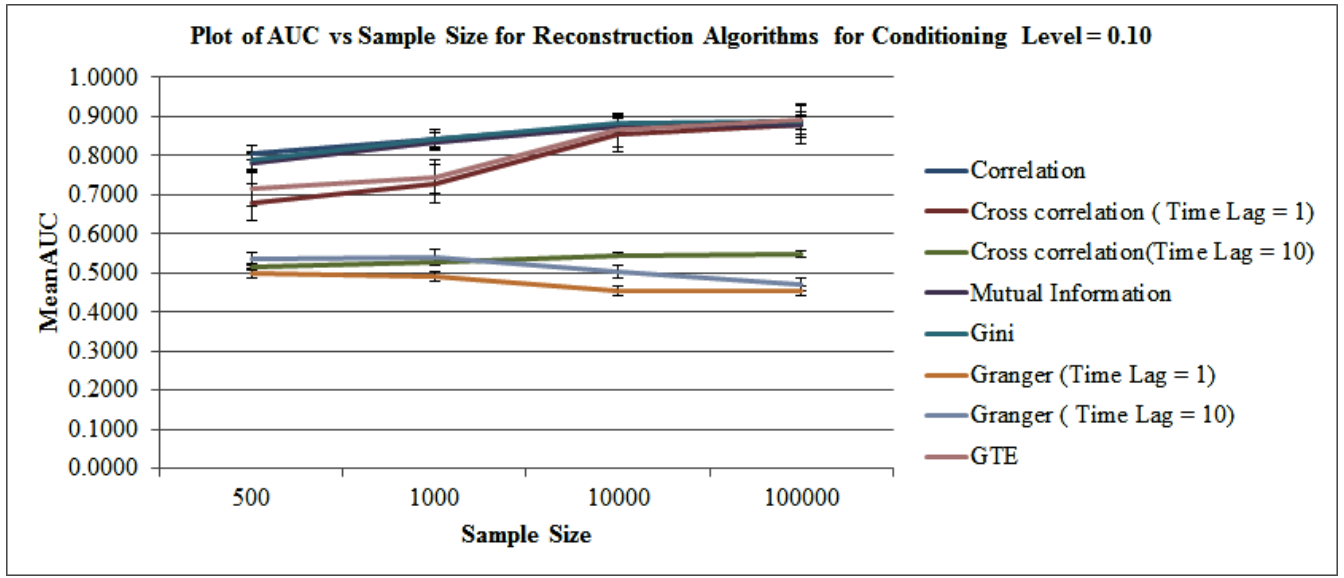

**FIG. 2.** Plot of AUC vs. Sample Sizes 500, 1000, 10000, and 100000 for Network 1 Conditioning level = 0.10.

We repeated the reconstruction procedure for smaller sample sizes from 100, 200… 1,000 for 1,000 runs and AUC and AUPR with 95% *confidence interval* (CI) were calculated. Results were averaged over all the repetitions for a given network and sample size. As can be seen from the ROC curve in Figure 3, correlation-based and information gain-based algorithms had AUC ~0.8 to ~0.9 for even small sample sizes of 1,000. As the lags were increased, both Cross-correlation and GC performed worse. This is explained as slower time lags fail to capture the characteristic instantaneous causal interactions between neurons.

The network reconstruction performance was evaluated with ensembles generated as described in Algorithm 1 for random samples of sizes 500, 1,000, 10,000, and 100,000 drawn from the original calcium fluorescence series and repeated for conditioning levels 0.10 and 0.15 for 1,000, 500, 100, and 10 runs, respectively with 95% C I.

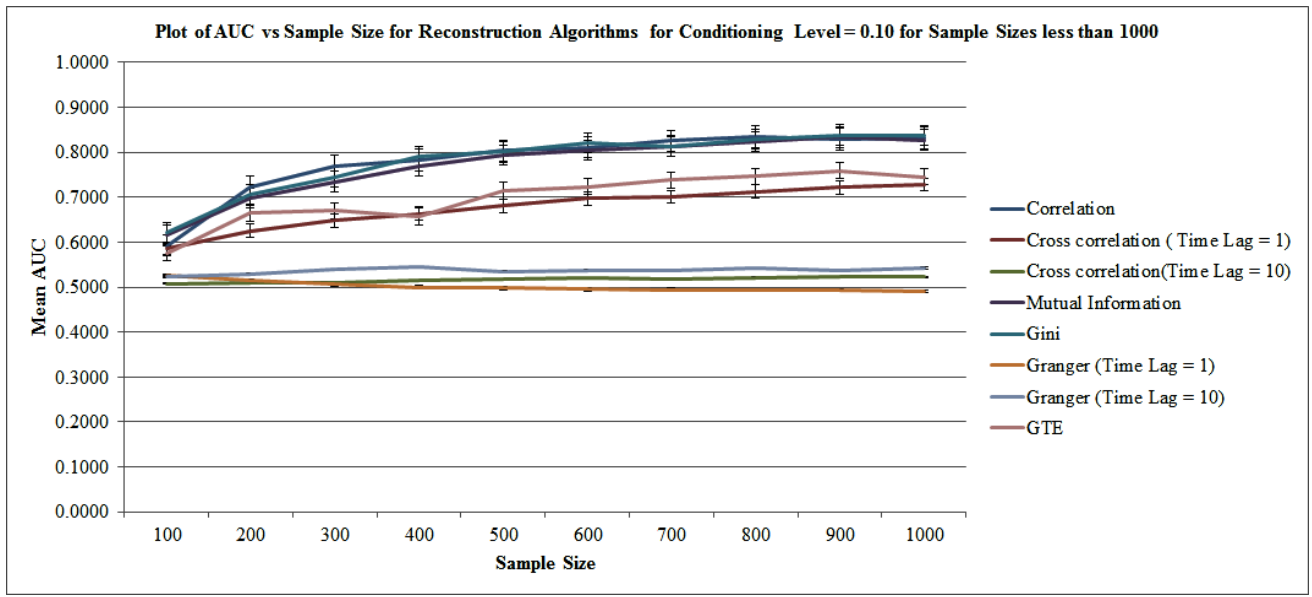

**FIG. 3.** Plot of AUC vs. Sample Sizes 100, 200… 1000 for Network 1 Conditioning level = 0.10.

The ensembles were generated using linear SVMs, NNs, and RFs. As can be observed in Figure 4 and Figure 5, the ensembles using RF and NN improved both AUC and AUPR performances over best-performing methods especially for larger sample sizes. RFs specifically generated an increase in AUC and AUPR even for sample sizes as small as 500.

These results suggest reliable reconstruction of AUC ~0.9 could be obtained even with smaller sample sizes of 10,000 compared to the total number of recorded samples of 180,000. These findings are important for experimental design of problems where recording large number of samples can be expensive and reconstruction using them time consuming.

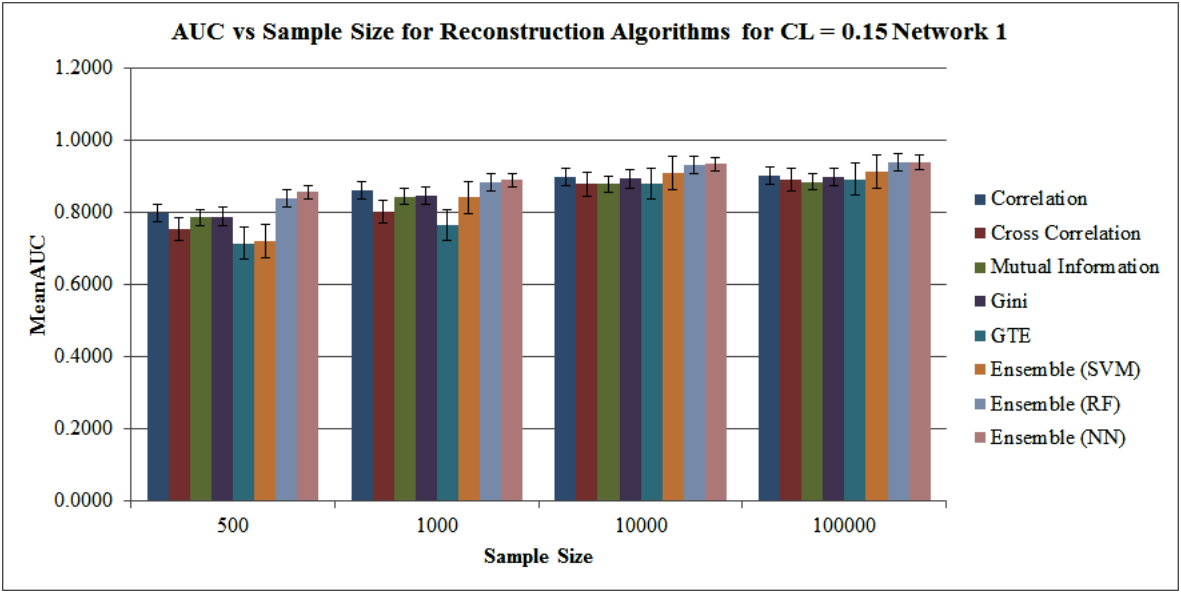

**FIG. 4.** Comparison of AUCs of reconstruction algorithms vs. Ensembles for Sample Sizes 500, 1000, 10000, and 100000 for Network 1 Conditioning level 0.15.

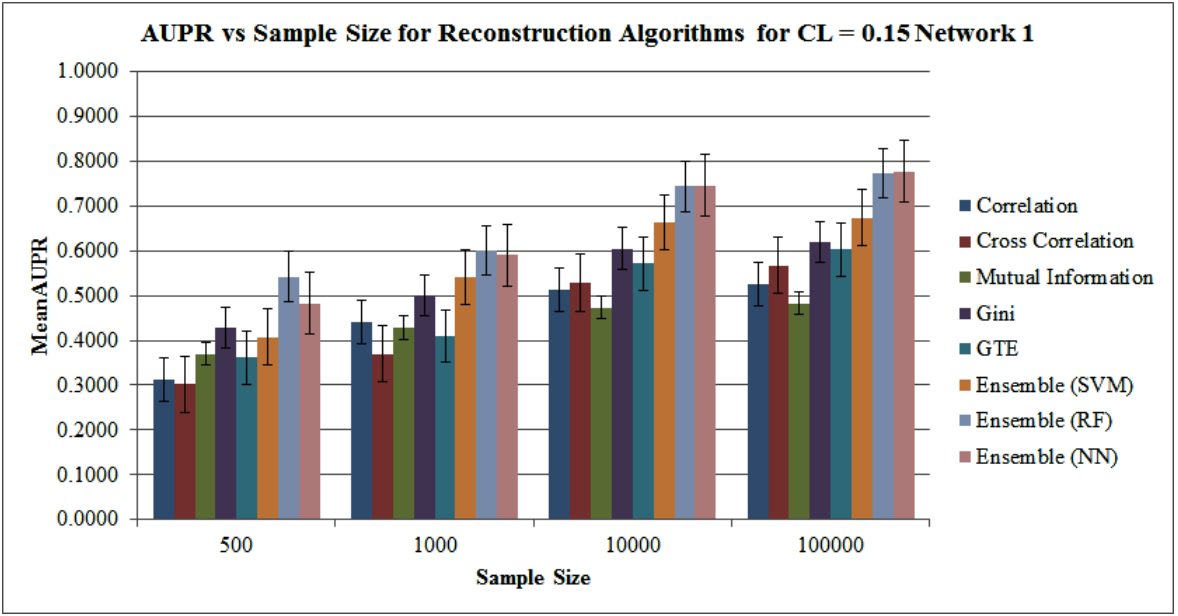

**FIG. 5.** Comparison of AUPRs of reconstruction algorithms vs. Ensembles for Sample Sizes 500, 1000, 10000, and 100000 for Network 1 Conditioning level 0.15.

## Discussion

The ability to accurately reconstruct models of brain connectivity is a critical step in understanding disease in a clinically useful manner. This is especially important when the connectivity is a root cause or intermediate pathologic process of symptoms or when the connectivity is a prognostic or diagnostic for clinical disease or response to treatment. It follows that additional studies that connect the relationship of connectivity with pathophysiology are needed for developing a complete picture of brain physiology and disease. We performed a benchmarking and ensembling study of several algorithms to reconstruct neuronal connectivity with different parameters based on sample size. Based on the performance of the algorithms, we used the scores on the connections generated from the best-performing base algorithms to construct features of a classifier and predict the presence or absence of connections. Our study advances the state of the art by adopting a principled, data-driven approach to feature construction which was informed by rigorous benchmarking based on sample size and parameters.

The net result is an important enhancement of reconstruction performance by about 3% of the best base predictors to RF and NN ensembles for large sample sizes of 100,000. Moreover, RF and NN ensembles generate AUCs ~0.9 with even 10000 samples. Ensemble, however, requires training data for the classifiers, unlike de-novo techniques, which do not require training.

This is a generalizable methodology applicable to other brain connectivity datasets with varying numbers of neurons. The methodology can be extended to incorporate signal preprocessing to filter out noise and infer directions of connections using edge orientation algorithms. Causal feature selection (Aliferis *et al.*, 2003) can be performed on the ensemble in future research. A variety of established model selection strategies can be used to choose optimal parameters, including nested cross validation approaches. When established negatives and positives exist for a dataset in hand, such methods can be used to tune parameters before applying to the unknown connections. When no prior knowledge exists for the system measured in a dataset, then such parameter optimization can be precede analysis by tuning to a number of similar datasets in the same or similar domains.

The limitations of this study include the use of simulated datasets as experimental data did not have the ground truth of the corresponding network. However, the simulator captures realistic sample sizes, light scattering artifacts, noise from nearby cells, etc. and is a faithful representation of experimental data. We focus in the present study on a simulated system which clearly allows systematic exploration of factors that may affect performance. As data from *in vivo* experiments become available with associated inferred networks, the methods described can be extended to real data. Our findings are extendible and generalizable as they can not only help advance the state-of-the-art methods to design scalable algorithms for modeling real neuronal systems and their dynamics but also aid causal structure reconstruction in other areas like finance, genomics, systems biology, and psychiatry where such problems are common.

**Conclusion**

In conclusion, our results towards reliably reconstructing neuronal networks from calcium fluorescence activity imaging data are very encouraging. The ensemble framework considerably improves both AUC and AUPR of existing individual techniques.

**Acknowledgement**

The first author thanks Dr. Isabelle Guyon, Javier Orlandi, Dr. Jordi Soriano, Dr. Olav Stetter, Dr. Demian Battaglia, and Dr. Mehreen Saeed for organizing the Kaggle Connectomics Challenge, providing sample code and data simulator as part of the Kaggle Connectomics Challenge, and useful discussions and advice regarding this research. The neuronal connectivity simulator and sample code in MATLAB used in the benchmarking experiments were obtained from the Kaggle Connectomics Challenge (Kaggle, 2014).

This work has utilized computing resources at the High Performance Computing Facility of the Center for Health Informatics and Bioinformatics at the NYU Langone Medical Center.

**Author Disclosure Statement**

No competing financial interests exist.